# STUDY OF THE STRUCTURE OF THE COMA CLUSTER BASED ON A HIERARCHICAL POWERFUL CLUSTERING METHOD

N. G. Kogoshvili,[1] T. M. Borchkhadze,[1] and A. T. Kalloghlian[2]

*Six subclusters in the Coma cluster have been selected on the basis of a hierarchical clustering method that takes the gravitational interaction among galaxies into account. Of these, 3 main subclusters around the galaxies NGC 4889, NGC 4874, and NGC 4839 have been singled out. We have used the objective statistical criterion applied by Vennik and Anosova in studies of close groups of galaxies to evaluate each member included in a subcluster with a high probability. Galaxies with a significant deficit of hydrogen HI, including objects from the Bravo-Alfaro list, have been identified with members of the subclusters, with the greatest number of them in the subclusters around NGC 4874 and NGC 4839. A quantitative estimate of the hydrogen deficit using the HI index in the RCG3 catalog reveals a statistically significant excess value for those galaxies that are members of the subclusters compared to the field galaxies with a hydrogen deficit in the studied area of Coma cluster. A substantial number of the spiral galaxies with a hydrogen deficit in the subclusters turned out to be radio galaxies as well.*

## 1. Introduction

The Coma cluster, which is one of the most well studied galaxy clusters because of its richness and nearly spherical shape, has long been regarded as a prototype for clusters in a state of dynamic equilibrium.

Various opinions have been expressed regarding the structure of this cluster. According to Kent and Gunn [1], the compact symmetric form of the Coma cluster is inconsistent with the existence of subclusters in it. West et al. [2,3] argue against the existence of subclusters in the central regions of clusters and view the Coma cluster, in particular, as a significant concentration of galaxies within a supercluster. Dressler and Shectman [4] remark that the probability of subclusters in Coma is <6%. Baier et al. [5] treat the Coma cluster as cD-like cluster with a single dominant NGC 4874

[1] E. K. Kharadze Abastumani Astrophysical Observatory, Georgia; e-mail: nmnt@yahoo.com
[2] V. A. Ambartsumyan Byurakan Astrophysical Observatory, Armenia; e-mail: astrofiz@sci.am

galaxy, based on data from the catalogue of Godwin et al. [6] and radio and x-ray data.

At the same time, a number of papers argue in support of the existence of subclusters in this cluster. Bahcall [7] first pointed out the nonuniform distribution of bright galaxies and suspected the existence of subclusters in the center of Coma. Rood [8] noted a tendency for S0 galaxies to group around NGC 4874 and for E galaxies, around NGC 4889. It has been found that the distribution of faint galaxies in Coma is more regular than that of bright galaxies, which, according to Biviano et al. [9], show a strong tendency toward clustering.

The first evidence of existence of subclusters in Coma was obtained from x-ray observations by the ROSAT satellite which revealed an irregular structure of Coma according to Jones et al. [10].

Various statistical methods for analyzing observational data have been used to detect the substructure in Coma. Fitchett and Webster [11], based on the Lee method, separated members of NGC 4889 group from NGC 4874 group in the centre of Coma. Escalera et al. [12] with the wavelet analysis selected two central subclusters in Coma cluster with 99% significance. Based on the S-tree method, Mazure and Gurzadyan [13] have studied the correlation between the parameters of a gravitationally interacting system of *N* bodies and revealed three subgroups within Coma. On the basis of a hierarchical clustering method, Shcherbanovskii [14] picked out 6 subclusters in the center of Coma, while noting the uncertainty in establishing their boundaries.

Most of these results require further confirmation, and the publication of new observations makes it possible to return anew to a discussion of these problems.

## 2. The method used for selection of subclusters in the Coma cluster

The method of hierarchical clustering proposed by Materne [15] is one of the various statistical tools used for selection of multiple groups of galaxies. This method was supplemented further by Tully [16] by the introduction of a special gravitational parameter that accounts the interaction among galaxies in selecting of physical groups.

This method was modified and used by Magtesyan [17] and then by Vennik and Anosova [18] for selecting of nearby groups of galaxies and comparing them with previously identified groups from other catalogs.

The criterion for selecting galaxies considered by Vennik and Anosova can be written in the form

$$F_{ij} = \max(M_i, M_j) r_{ij}^{-2}, \quad i, j = 1, \ldots, N \quad i \neq j, \tag{1}$$

where $r_{ij}$ is the distance between the two galaxies with masses $M_i$ and $M_j$. Galaxies are organized into a group based on optimization of this parameter.

We have used the ideology of the method employed by Vennik and Anosova for selecting subclusters in the Coma cluster. In order to exclude the influence of possible members of Coma Supercluster and galaxies in the fore- and background, we have restricted the region of study for the Coma cluster and its nearest surroundings to the ranges $\alpha = 12^h 30^m \div 13^h 30^m$, $\delta = 26^\circ \div 31^\circ$, and $V = 5300 \div 9000$ km/s, which are close to the conventional ranges according to Gavazzi et al.[19]). Galaxies for this region were chosen from the Merged catalogue of galaxies compiled by Kogoshvili and Borchkhadze [20] on the basis of data from a majority of catalogues of bright galaxies. 205 galaxies were included in the sample, most of them brighter than $15^m.5$ with weighted mean values of radial velocities corrected with respect to the GSR from the RCG3 catalogue of de Vaucouleurs et al.[21], included as well in the Merged catalogue. Figure 1 shows the 2-D coordinate distribution of these galaxies. Taking into account that the mass of a galaxy is proportional to its luminosity, the selection criterion was written in the form

$$F_{ij} = \max(L_i, L_j)/V_\sigma \quad \text{where} \quad i = 1,...,k; \quad j = 1,...,n; \quad N = n+k; \tag{2}$$

$$V_\sigma = \frac{4}{3H^3}\left(V_i^3 - V_j^3\right) \text{tg} \frac{\delta_i - \delta_j}{2} \text{tg} \frac{(\alpha_i - \alpha_j)\cos(\delta_i + \delta_j)/2}{2}, \tag{3}$$

where $V_\sigma$ is the volume occupied by two galaxies which satisfy the selection criterion.

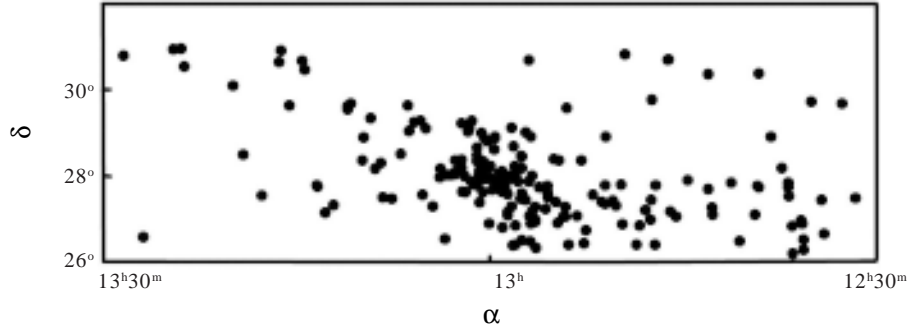

Fig. 1. The distribution of 205 galaxies within the confines of the region of the Coma cluster studied here.

It should be noted, however, that the hierarchical clustering method is based on the selection of a pair of galaxies and replacing it with a hypothetical object for subsequent selection of new pairs until completion of the structure, which frequently tends to yield elongated formations.

We have chosen a way, selecting several of the brightest galaxies in Coma and calculating the values $F_{ij}$ separately for each of them with every galaxy from the remaining galaxies in the sample. Galaxies, in accordance with the basic condition $F_{ij} = F_{max}$, are selected into a subcluster taking into account the maximum value of a gravitational interaction between two galaxies occupying a minimum volume. The merging ptocess continues until all the objects join subclusters. When the same object is found in several substructures, we consider the highest value of $F_{ij}$ as the preference for inclusion of a galaxy into appropriate subgroup. The major difficulty with the hierarchical clustering method is the absence of an objective criterion for establishing the boundaries of selected substructures.

Anosova [22] has proposed an objective statistical criterion which makes it possible to identify multiple groups of galaxies with physical connection of components and to select nonrandom groups. This method has been used previously to select subclusters in the Virgo cluster [23].

According to this method, the mathematical expectation $E_n$ of the number of random groups with $n$ objects occupying volume $V_\sigma$ in a sample of $N$ objects with volume $V_\Sigma$ may have the form

$$E_n = C_N^n B^{n-1}[1-B]^{N-1}, \tag{4}$$

where $B = V_\sigma/V_\Sigma$, $V_\sigma$ is determined using Eq. (3), and

$$V_\Sigma = \frac{4}{3H^3}\left(V_{max}^3 - V_{min}^3\right) \text{tg} \frac{\delta_{max} - \delta_{min}}{2} \text{tg} \frac{(\alpha_{max} - \alpha_{min})\cos(\delta_{max} + \delta_{min})/2}{2}. \quad (5)$$

Vennik and Anosova have supplemented this method by introducing a parameter which would make it possible to estimate the probability of including physical members in a group. Since we are mainly considering cases with $n = 2$, in fact $E_2 \equiv E$ and the new parameter takes the form $K = N/2E$ with the limits for $K \approx 10$ and $\log K_{lim} \approx 1$. The following condition must be satisfied:

$$\text{if} \quad E < 1 \quad \text{and} \quad K_n > K_{lim},$$

then each object which satisfies this condition can be regarded as a physical member of the selected group with high probability. Otherwise, the corresponding object is classified as random.

## 3. The selection of subclusters in the Coma cluster

Based on the hierarchical clustering method with the gravitational interaction among galaxies taken into account, six subclusters were selected in the Coma cluster. Each galaxy was included in a subcluster in accordance with the computed value of the statistical parameter $\log K_n$ within the limits of restrictions given above.

Two subclusters were selected in the center of the Coma cluster around the galaxies NGC 4889 and NGC 4874, one subcluster in the SW condensation of galaxies around NGC 4839, and three smaller substructures around the galaxies NGC 4816, NGC 4789, and NGC 4555.

For galaxies in three main subclusters Table 1 lists the following: the number in the Paturel et al. catalogue [24], the NGC/IC number, the morphological classification of a galaxy according to the RCG3 catalogue, and the calculated value of the statistical parameter $\log K_n$. A supernova observation is indicated by an asterisk *. A hydrogen HI deficit in a galaxy is indicated by the symbol HI and an identification with radio galaxies from the Venturi et al. catalogue [25] by the symbol R. Table 2 lists data for the other three substructures.

TABLE 1. List of Galaxies Contained in the three Main Subclusters in the Coma Cluster

| Subcluster I | | | | Subcluster II | | | | Subcluster III | | | |
|---|---|---|---|---|---|---|---|---|---|---|---|
| PGC | NGC/IC | Type | log$k$ | PGC | NGC/IC | Type | log$k$ | PGC | NGC/IC | Type | log$k$ |
| 1 | 2 | 3 | 4 | 5 | 6 | 7 | 8 | 9 | 10 | 11 | 12 |
| P44715 | N4889 | E | | *P44628 | N4874 | E | R | P44298 | N4839 | E | R |
| P44698 | N4886 | E | 4.89 | P44624 | N4872 | L | 5.74 | P44337 | N4842A | E | 4.71 |
| P44566 | N4864 | E | 4.07 | P44606 | N4871 | L | 4.55 | P44338 | N4842B | E | 4.22 |
| P44736 | N4898 | E | 4.00 | P44621 | N4873 | L | 3.64 | P44268 | | S? | 4.15 |
| P45140 | I 4133 | E? | 3.35 | P44658 | N4876 | E | 3.20 | P45526 | | S | 3.24 |
| P44726 | I 4021 | E | 3.15 | P44467 | | L? | 3.16 | P44437 | | S HI | 2.55 |
| P44804 | I 4041 | E | 3.00 | P44587 | N4869 | E | 3.16 R | P44037 | N4807 | L | 2.39 |
| P44449 | N4850 | L | 2.86 | P44553 | I3959 | L? | 2.70 | P44722 | | L? | 2.39 |

| 1 | 2 | 3 | 4 | 5 | 6 | 7 | 8 | 9 | 10 | 11 | 12 |
|---|---|---|---|---|---|---|---|---|---|---|---|
| P44686 | N4881 | E | 2.75 | P44633 | I3990 | S? | 2.50 | P44324 | N4840 | E | 2.38 |
| P45027 | N4929 | E | 2.35 | *P44968 | N4926A | L HI | 2.44 | P44322 | I 837 | S?HI | 2.37 R |
| P44554 | I 3957 | L? | 2.32 | P44885 | N4919 | L | 2.44 | P41468 | I 3454 | S | 2.33 |
| P44818 | I 4045 | E | 2.26 | P44575 |  | E | 2.38 | P44319 |  | S? | 2.29 |
| P44176 | N4828 | S? | 2.16 | P44632 | I3991 | S | 2.38 | P44481 | N4853 | L | 2.25 |
| P44697 | N4892 | S | 2.11 | P44405 | N4848 | S HI | 2.32 R | P44263 |  | S? | 2.20 |
| P43455 | N4728 | E | 2.10 | P44524 | I3949 | L HI | 2.29 R | P44147 | I 3913 | S?HI | 2.20 |
| P44789 | I 4040 | S HI | 1.99 R | P44551 | I3960 | L | 2.29 | P43387 | A1246B | E | 2.09 |
| P45082 | N4934 | S? | 1.98 | P44795 | I842 | S?HI | 2.15 | P44779 |  | S?HI | 2.05 R |
| P44819 | N4907 | S HI | 1.93 | *P44068 | I3900 | L | 2.07 | P44178 | N4827 | E/L | 2.00 R |
| P44756 | I 4032 | E | 1.93 | P44945 | N4927 | L | 2.01 R | P43981 | N4798 | L | 1.87 |
| P44864 |  | E | 1.89 | P43930 |  | S | 1.93 | P44323 | N4841A | E | 1.83 |
| P44768 | N4896 | L | 1.87 | P44534 | N4859 | S | 1.87 | P42083 | I 3587 | S | 1.83 |
| P44737 | N4895 | L | 1.81 | P44539 | N4860 | E | 1.85 | P43539 | N4745 | L? | 1.79 |
| P45055 | N4931 | L | 1.80 | P44667 |  | S | 1.83 | P45471 |  | S | 1.67 |
| *P44899 | N4921 | S HI | 1.76 R | P44647 |  | S?HI | 1.82 | P44225 |  | E? | 1.61 |
| P44508 | I3946 | L | 1.77 | P45271 |  | S | 1.81 | P44486 |  | L HI | 1.61 |
| P42934 |  | S | 1.74 | P44921 | I4088 | S HI | 1.67 | *P43514 | A1249 | E | 1.56 |
| P44828 | N4908 | E | 1.62 | P45025 | I4106 | S | 1.54 | P44329 | N4841B | E | 1.46 |
| P45997 | N5032 | S HI | 1.54 | P44896 | N4922 | IrHI | 1.53 | P42067 | I 3585 | L | 1.46 |
| P44822 |  | S | 1.53 | *P45133 | N4944 | S | 1.46 | P43139 |  | Ir? | 1.41 |
| P42721 |  | S | 1.52 | P45668 |  | S | 1.43 | P44200 | I 835 | S? | 1.36 |
| P43509 | N4735 | S?HI | 1.34 | P44840 | N4911 | S HI | 1.40 R | P43995 |  | E | 1.32 |
| P45406 | N4971 | L? | 1.26 | P44541 | Mr | S HI | 1.39 R | P44416 |  | S?HI | 1.31 R |
| P45542 | N4983 | S? | 1.18 | P43726 |  | S?HI | 1.34 R | P45253 | N4957 | E | 1.28 |
|  |  |  |  | P44848 |  | E? | 1.18 | P43164 |  | S? | 1.23 |
|  |  |  |  | P44973 |  | S? | 1.03 | P45023 |  | L? | 1.23 |
|  |  |  |  |  |  |  |  | *P44938 | N4926 | L | 1.20 |
|  |  |  |  |  |  |  |  | P44144 | N4819 | S | 1.19 |
|  |  |  |  |  |  |  |  | P44196 |  | L | 1.18 |
|  |  |  |  |  |  |  |  | P41980 | N4556 | E | 1.12 |
|  |  |  |  |  |  |  |  | P42765 |  | Ir?HI | 1.08 |
|  |  |  |  |  |  |  |  | P46302 |  | S HI | 1.07 |
|  |  |  |  |  |  |  |  | P44908 | I 843 | L | 1.03 |

The galaxies in the subclusters are arranged according to their insertion in the group per the criterion $\log K_n$ in a direction from the center outward. The dependence of $\log K_n$ on $n/n_{gr}$, for the three central subclusters, where $n/n_{gr}$ is a number of each galaxy in the corresponding subcluster, is approximated by $\log K_n = (3.6 \pm 0.20) - (2.7 \pm 0.02) n/n_{gr}$.

TABLE 2. Galaxies Contained in Subclusters IV, V, and VI of the Coma Cluster

| PGC | NGC/IC | Type | log k | PGC | NGC/IC | Type | log k | PGC | NGC/IC | Type | log k |
|---|---|---|---|---|---|---|---|---|---|---|---|
| P44114 | N4816 | L | | P43895 | N4789 | E/L | R | P41975 | N4555 | E | |
| P44552 | I 3960A | E? | 3.31 | P43875 | N4787 | S | 3.68 | P41974 | | S | 4.71 |
| P44567 | I 3963 | L | 2.71 | P43773 | | S? | 2.35 | P42479 | I 3645 | S? | 2.39 |
| P44148 | N4821 | E | 2.55 | P44487 | | S | 1.93 | P45097 | | S?HI | 2.37 |
| *P44364 | | E | 2.55 | P43618 | | E? | 1.80 | P42331 | | L? | 2.10 |
| P44043 | | L? | 2.53 R | P43874 | N4788 | S? | 1.78 | P43142 | | S | 1.78 |
| P44044 | | S | 2.50 | P43008 | N4673 | E | 1.66 | P41995 | N4557? | S? | 1.69 |
| P43399 | N4715 | L | 2.50 | P43511 | | S? | 1.66 | P42060 | I 3582 | S HI | 1.67 |
| P44212 | | | 2.50 | P43952 | | E | 1.55 | P41774 | I 3508 | L? | 1.30 |
| P44850 | | S? | 2.43 | P44502 | N4854 | L | 1.37 | P43708 | I 831 | E | 1.16 |
| P44151 | | L | 2.35 | P43686 | | S?HI | 1.28 | P43278 | | S | 1.00 |
| P43256 | A1246A | S | 1.98 | P43848 | I 832 | E? | 1.15 | | | | |
| P42098 | I 3593 | S? | 1.93 | P43200 | N4692 | E | 1.07 | | | | |
| P45890 | | S | 1.73 | *P44386 | | S | 1.07 | | | | |
| P44138 | I 834 | S | 1.49 | P43437 | N4721 | L? | 1.02 | | | | |
| P41808 | I 3516 | | 1.45 | | | | | | | | |
| P45940 | | S? | 1.31 | | | | | | | | |
| P42314 | | E? | 1.26 | | | | | | | | |

According to Table 1, galaxies of different morphological types have different radial distributions within the selected subclusters. A noticeable clump of elliptical galaxies is observed in the central region of the I subcluster around NGC 4889. A greater concentration of lenticular galaxies compared to ellipticals is seen in the II subcluster around NGC 4874. It is evident as well a more scattered distribution of spirals in the outer regions of both subclusters. The SW subcluster around NGC 4839 is distinguished by a mixed distribution of all types of galaxies.

TABLE 3. Mean Values of Radial Velocities and Absolute Magnitudes of Galaxies in the Inner and Outer Parts of three main Subclusters

| | Subcluster I | | | Subcluster II | | | Subcluster III | | |
|---|---|---|---|---|---|---|---|---|---|
| | $<V>$ | $<M>$ | $n_M$ | $<V>$ | $<M>$ | $n_M$ | $<V>$ | $<M>$ | $n_M$ |
| Inner part | 6427 | -19.6 | 11E+2L+2S | 6951 | -19.9 | 5E+10L+4S | 7283 | -20.0 | 5E+6L+9S |
| σ | 362 | 0.68 | | 521 | 0.51 | | 381 | 0.67 | |
| Outer part | 6601 | -19.8 | 3E+2L+10S | 7141 | -20.1 | 2E+13S+1Ir | 7226 | -19.8 | 6E+7L+7S+2Ir |
| σ | 885 | 0.71 | | 737 | 0.80 | | 460 | 0.81 | |

TABLE 4. Parameters of Radial Velocities of Galaxies for 6 Selected Subclusters in the Coma Cluster

|  | Subcluster I | Subcluster II | Subcluster III | Subcluster IV | Subcluster V | Subcluster VI |
|---|---|---|---|---|---|---|
| $n$ | 33 | 35 | 42 | 18 | 15 | 11 |
| $\langle V \rangle$ | 6522 | 7038 | 7253 | 6948 | 7538 | 6911 |
| $\sigma$ | 714 | 646 | 430 | 443 | 768 | 516 |
| $M_3$ | 1.7 | -1.0 | -1.4 | 0.8 | -0.9 | 1.1 |
| $M_4$ | 3.6 | 1.2 | 1.6 | 1.8 | -0.8 | 0.1 |

Table 3 lists mean values of radial velocities and absolute magnitudes of galaxies for the inner and outer regions of three main subclusters. According to Tables 1 and 3 morphological segregation is evident for galaxies in the I and II subclusters with a certain increase in mean values of velocities and, especially, in their dispersion in the outer regions. There is no evidence of luminosioty segregation for the types of bright galaxies considered here. Conserning the SW subcluster, no morphological, luminosity or velocity segregation is observed in it.

Substructures IV, V, and VI can be regarded as multiple groups because of the small number of galaxies in them.

Table 4 lists mean values of radial velocity parameters of galaxies computed for 6 selected subclusters where $n$ is the number of galaxies, $\langle V \rangle$ - the average velocity, $\sigma$ - the standard deviation, $M_3$ - the asymmetry, and $M_4$ - the excess. Because some overlap was observed in the distribution of radial velocities of galaxies in subclusters I and II, the velocity distributions in these subclusters were compared using the t-test assuming that the null hypothesis of equality of the average values of these distributions is true. The resulting $\alpha = 0.0028$ allowed us to reject the null hypothesis with a high probability $P = 1 - \alpha$ and treat these subclusters as independent, spatially separated groups. Histograms of the velocities of subclusters I and II are shown in Fig. 2 and the 3-D spatial distributions of the galaxies in the 6 selected galaxy subclusters are shown in Figs. 3 and 4.

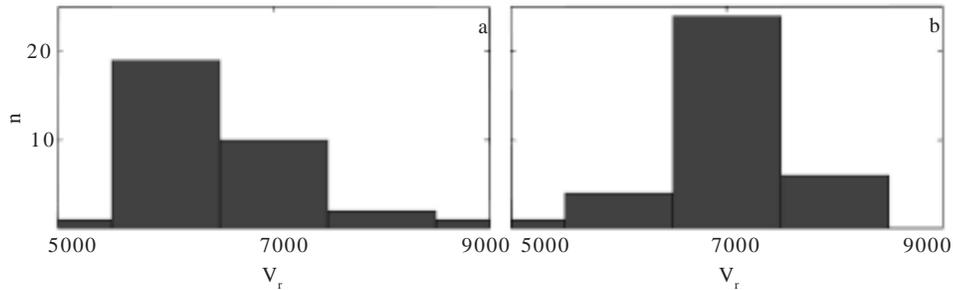

Fig. 2. Histograms of radial velocities of galaxies in (a) subcluster I around NGC 4889 and (b) subcluster II around NGC 4874.

## 4. Identifying galaxies with a hydrogen HI deficit in the selected subclusters

According to Gavazzi et al. [26], the Coma cluster stands out among other clusters in having a substantial hydrogen HI deficit. Based on this criterion, Gavazzi concludes [27] that most of the spiral galaxies seen in this region are true members of the cluster.

Bravo-Alfaro et al. [28] have obtained with the VLA radio telescope images of 19 spiral galaxies at 21 cm which indicate a noticeable hydrogen HI deficit in them. All of these galaxies were found in the cental 30 ' region of

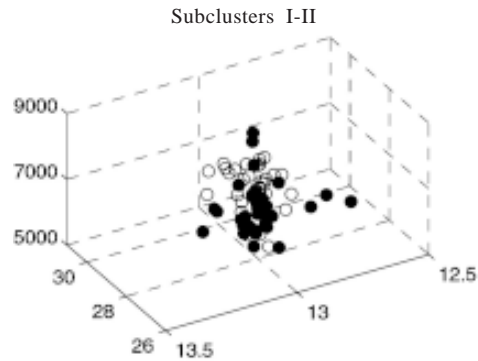

Fig. 3. The 3-D spatial distributions of galaxies in subcluster I around NGC 4889 (solid dots) and subcluster II around NGC 4874 (hollow circles).

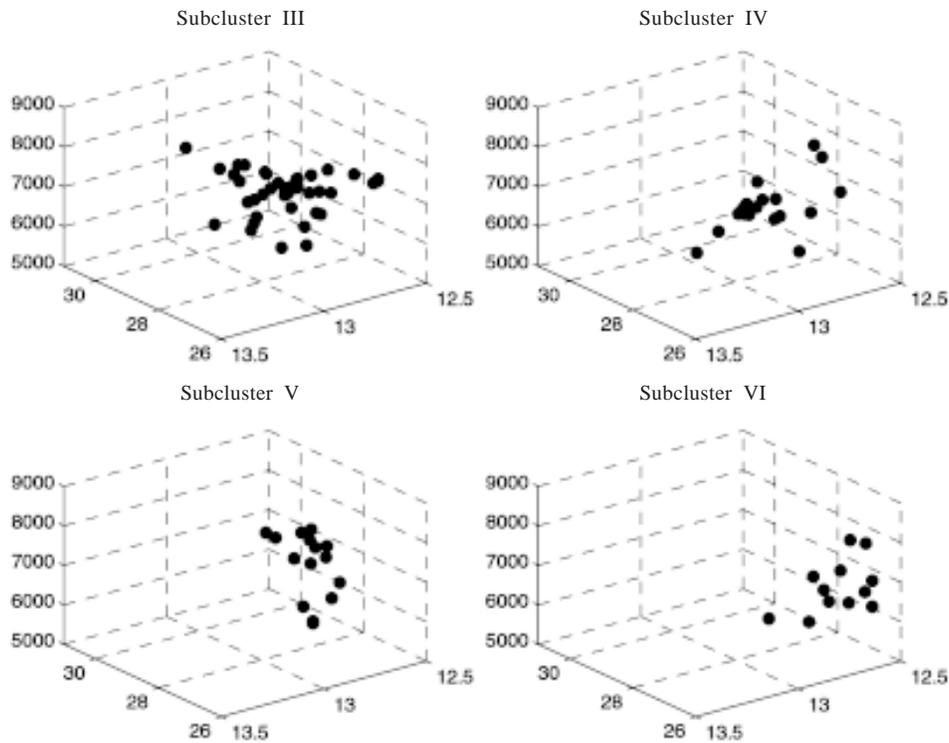

Fig. 4. The 3-D spatial distributions of galaxies in the subclusters around NGC 4839 (III), NGC 4816 (IV), NGC 4789 (V), and NGC 4555 (VI).

Coma cluster. 16 of them turned out to be the members of 3 main subclusters with the largest number of galaxies (9) - in the subcluster II around NGC 4874, 5 - in the III, and 2- in the I. According to Bravo-Alfaro's evidence, 7 of these galaxies show the perturbed distribution of hydrogen with sizes smaller than their optical disks and often not coinciding with them. 4 of these objects were identified with spirals in the outer region of the II subcluster around NGC 4874, surrounding the denser distribution of E and L galaxies, 2 were found in the I, and 1-in the III subcluster.

The RCG3 catalog gives values for the hydrogen HI index determined by analogy with the color index $HI = m_{21}^0 - B_T^0$. We have verified all spiral galaxies in the studied area of Coma cluster on the HI index. As it turned out only 10 objects from Bravo-Alfaro's list of 18 galaxies had HI index measurements, but we found the other 8 new galaxies with HI index among late type members in selected subclusters. They all are indicated in Table 1 by symbol HI. In addition, 22 galaxies with measured HI indices were identified outside the subclusters in the studied area of Coma. The galaxies with measured $(B-V)_T^0$ color in RCG3 were selected from among the galaxies with an HI index. Table 5 presents the average values of HI indices, $(B-V)_T^0$ colors and radial velocities for spirals in the subclusters and in the general field of the cluster. A comparison of the mean values of HI indices for these two groups of galaxies yielded a statistically significant difference at a level of $P = 1 - \alpha$, where $\alpha = 0.0035$, confirming that spiral galaxies identified in subclusters evidence larger HI deficit than spirals in the field.

The hydrogen HI deficit in spiral galaxies associated with subclusters, distortion in their HI distribution, that sometimes does not coincide with optical disk of a galaxy, reflects the effect of interaction of galaxies with intercluster medium during their motion in the cluster, that results in sweeping of HI hydrogen out of them by dynamical pressure.

A comparison of the $(B-V)_T^0$ color values obtained for the galaxies in both groups in Table 5 suggests that most of the galaxies with a hydrogen deficit are young blue objects. If we count the $(B-V)_T^0$ color separately just for the galaxies in subclusters II and III, then we obtain $(B-V)_T^0 = 0.62$ for 8 galaxies. According to Bravo-Alfaro, the interaction of these galaxies with intercluster medium could trigger a burst of star formation in these galaxies.

Based on the ROSAT x-ray survey and dynamical analysis of available radial velocities of galaxies Colless [29] emphasizes that Coma cluster cannot be considered as dynamically relaxed system and primary attention has to be shifted to studies of irregular processes and history of the formation of this cluster.

TABLE 5. Average Values of the Index $HI = (m_{21}^0 - B_T^0)$, $(B-V)_T^0$ Color, and Radial Velocity for Galaxies with an HI Deficit within and Outside the Selected Subclusters

| Subclusters | | | | | Field of Coma cluster | | | | HI index Excess |
|---|---|---|---|---|---|---|---|---|---|
| $n$ | HI | V | $n$ | $(B-V)_T^0$ | $n$ | HI | V | $n$ | $(B-V)_T^0$ | |
| 18 | 2.66 | 6941 | 13 | 0.67 | 22 | 1.83 | 6879 | 7 | 0.65 | 0.83 |
| $\sigma$ | 0.26 | 192 | | 0.04 | | 0.12 | 195 | | 0.11 | 0.29 |

In Table 1, among the selected subclusters we have indicated supernovae observed in galaxies. The largest number of them was marked in subcluster II around NGC 4874. Supernovae are responsible for injecting relativistic electrons which facilitate heating of intergalactic medium as they interact with galactic magnetic fields.

The Coma cluster is anomalous in its radio and x-ray emission. Venturi et al. [29] have identified 29 radio sources, detected at a frequency of 326 MHz, in galaxies in the region of the Coma cluster, with a reliable radio flux taken to involve a level $S_{326} \geq 5$ mJy. 20 of the radiogalaxies in this list were in the region of Coma that we have studied and 17 of them were identified as members of the selected subclusters. Table 6 lists the average values of the radio flux $\log S_{326}$ and absolute stellar magnitude for the radio galaxies in these subclusters as functions of their morphological type. Tables 1 and 6 show that the high-luminosity elliptical galaxies in the center of subclusters II, III, and V were strong sources of radio emission, while in the condensation of E galaxies in subclusterI around NGC 4889 no radio source with $S_{326} \geq 5$ was observed. In addition, no difference in the radio emission from the S and L galaxies in the selected groups or dependence of the radio emission on their luminosity was observed. At the same time, a significant number of galaxies with a hydrogen HI deficit turned out to be radio galaxies. The Coma cluster is permeated by strong diffuse radio emission with the central radio source C, that testifies the existence of relativistic electrons and large-scale magnetic fields, but some mechanism is required to reaccelerate their energy.

According to theorists, passage of a group of galaxies through the center of a cluster in cosmologically not too distant past could be one such energy source.

The large number of galaxies with an HI deficit, showing sometimes distortion in its distribution, and identified with the members of subcluster II around NGC 4874, makes it possible to suppose that this subcluster is passing through the Coma cluster with ongoing merging with the subcluster I around NGC 4889.

According to Bravo-Alfaro none of the galaxies near NGC 4839 has been observed at 21 cm line.

We have found 8 galaxies with a significant HI deficit in the subcluster around NGC 4839. Of these, 5 are in Bravo-Alfaro's list, and one of them has a distorted hydrogen structure. This fact, along with the existence of a radio tail in the radio galaxy NGC 4839 directed away from the center of Coma, may evidence of movement of this subcluster III with NGC 4839 toward the center of Coma cluster at the present time.

TABLE 6. Average Values of the Radio Flux $\log S_{326}$ and Absolute Stellar Magnitude of Radio Galaxies of Different Morphological Types in the Selected Subclusters

|  | E | L | S |
|---|---|---|---|
| $\langle \log S_{326} \rangle$ | 2.68 | 1.22 | 1.41 |
|  | 0.35 | 0.23 | 0.33 |
| $\langle M \rangle$ | -21.1 | -19.9 | -20.1 |
|  | 0.33 | 0.31 | 0.28 |
| $n$ | 5 | 3 | 9 |

## 5. Conclusions

Based on the method of hierarchical clustering and taking the gravitational interaction among galaxies into account 6 subclusters were selected in the Coma cluster. Of these three main subclusters were singled out around the galaxies NGC 4889, NGC 4874, and NGC 4839. An objective statistical criterion developed by Vennik and Anosova has been used to estimate every physical member included in a subcluster with a high probability.

Galaxies of different morphological types displayed different radial distributions in the selected subclusters. A noticeable clumping of E galaxies in the central region of the subcluster around NGC 4889 has been confirmed, along with a higher concentration of L type compared to E type galaxies in the subcluster around NGC4874. The more extended distribution of S galaxies in the outer regions of both subclusters is noted. The subcluster around NGC 4839 shows a mixed distribution of all types of galaxies.

Galaxies with a significant hydrogen HI deficit, including objects from the list of Bravo-Alfaro, have been identified with members of main subclusters with their greater number in the subclusters around NGC 4874 and NGC 4839. A quantitative estimation of the hydrogen deficit by the HI index in the RCG3 catalog revealed a statistically significant exceeding of its value for galaxies belonging to subclusters compared to galaxies in the field of Coma cluster. A substantial number of spiral galaxies with a hydrogen HI deficit found in subclusters turned out to be radio galaxies, as well.

The low hydrogen HI content in spiral galaxies associated with subclusters in the Coma cluster with its sometimes distorted distribution reflects the effect of interaction of galaxies with the intercluster medium during possible passage of subclusters through the Coma cluster.

This all suggests that the subclusters themselves could be considered as an area of active processes in galaxy clusters.